\documentclass[aps,prb,twocolumn,showpacs]{revtex4}

\usepackage{graphicx}

\newcommand{\thub}{{\tilde t}}
\newcommand{\Uhub}{{\tilde U}}

\begin{document}

\title{Localization of two interacting electrons in quantum dot 
arrays driven by an ac-field}
\author{C.~E.~Creffield}
\affiliation{Dipto di Fisica, Universit\`a di Roma
``La Sapienza'', Piazzale Aldo Moro 2, I-00185 Rome, Italy}
\author{G.~Platero}
\affiliation{Instituto de Ciencia de Materiales (CSIC), Cantoblanco, E-28049
Madrid, Spain}

\date{\today}

\begin{abstract}
We investigate the dynamics of two interacting electrons moving in a
one-dimensional array of quantum dots under the influence of an ac-field. 
We show that the system exhibits two distinct regimes of behavior, depending
on the ratio of the strength of the driving field to the inter-electron
Coulomb repulsion. When the ac-field dominates, an effect termed
coherent destruction of tunneling occurs at certain frequencies, in
which transport along the array is suppressed. In the other, weak-driving,
regime we find the surprising result that the two electrons can
bind into a single composite particle -- despite the strong Coulomb
repulsion between them -- which can then be controlled by the
ac-field in an analogous way. We show how calculation of the
Floquet quasienergies of the system explains these results, and thus
how ac-fields can be used to control the localization 
of interacting electron systems.
\end{abstract}

\pacs{73.23.-b, 72.15.Rn, 03.67.Mn, 42.50.Hz}
 
\maketitle

\section{Introduction}

Recent years have seen considerable progress in the theoretical
and experimental investigation of electron dynamics in mesoscopic systems.
Preservation of the quantum coherence of the electronic states can
give rise to unusual transport phenomena, one of the most
notable of which is termed ``coherent destruction of tunneling''
\cite{hanggi_prl,hanggi_epl} (CDT), or alternatively 
``dynamical localization''. This occurs when the tunneling of a particle
between potential minima is suppressed, 
thus localizing the particle, under the influence of an
oscillatory electric field. The degree of suppression depends sensitively
on the parameters of the driving field, thereby giving the attractive
possibility of coherently manipulating electronic states in semiconductor
nanostructures by means of laser pulses or oscillating gate potentials.

The presence of CDT was first recognized in quantum well systems
\cite{hanggi_prl,hanggi_epl} and superlattices 
\cite{dunlap, holthaus, zhao}, in which the effects of inter-electron 
interactions are negligible due to the high concentration of carriers.
More recent work \cite{tamb_met,zhang1,zhang2,pasp}
has focused on quantum dot (QD) systems, which give
the ability to study a small number of electrons in a highly-controllable
environment. In such nanostructures, however, the low electron-density
means that correlations produced by the Coulomb interaction cannot be 
neglected, and indeed can have extremely significant effects 
\cite{cecgp1,cecgp2}.
Numerical investigations of a system of two electrons in a double QD
have revealed novel forms of CDT, resulting from the inter-electron 
Coulomb interaction, which have been successfully analyzed by 
means of the Floquet approach \cite{cecgp1}.
Here we extend and generalize these results by investigating the
case of two electrons confined to a one-dimensional {\em array} of
QDs, and make a detailed study of
the interplay between the driving field and the Coulomb interaction
on coherent transport. Although interesting in its own right,
such understanding is also of practical importance to the field of
quantum computation \cite{loss}, as this type of structure provides a 
promising method of implementing scalable arrays of quantum bits
\cite{elzerman}.

In this investigation we first demonstrate that the Coulomb interaction
produces two regimes of behavior, namely when the driving is
stronger or weaker than the Coulomb interaction. This was seen previously
in the double QD system \cite{cecgp1}, and we show that it is
a generic feature, common to all sizes of array. In the strong
driving regime CDT occurs when the field frequency is in resonance
with the Coulomb energy at certain field strengths which can be predicted
by Floquet theory \cite{cecgp1}. In the weak-driving regime, however,
no resonance effects arise, and initial states in which a QD is 
doubly-occupied show a surprising behavior \cite{pasp,cecgp3,noba} 
in that despite the strong Coulomb repulsion between them, the two
electrons can bind together to form a single composite particle of
charge $2e$. The localization and tunneling dynamics of this two-electron 
state can then be controlled by suitably choosing the parameters of 
the ac-field, in a similar way to the control of a single particle.

\section{Model and methods}

We consider a system of two-electrons confined to a one-dimensional array
of QDs, with a time-dependent electric field $F(t)$ applied parallel to the 
array. Such an array can be produced experimentally by, for example, 
gating a two-dimensional electron gas confined in a semiconductor 
heterostructure \cite{elzerman, waugh}. We model the QD array as a 
single-band tight-binding model of Hubbard type. For the double QD case, 
explicit calculation has shown that such an effective model is able to 
reproduce well the behavior of a more realistic model \cite{cecgp1},
and accordingly for an array of $N$ QDs we use the effective 
lattice Hamiltonian:
\begin{equation}
H = - \thub \sum_{\langle i,j \rangle, \sigma}^N
\left[ c^{\dagger}_{i \sigma} c^{ }_{j \sigma} + H.c. \right]
+ \sum_{j=1}^N  \left[ \Uhub n_{j \uparrow} n_{j \downarrow} 
+ e a F(t) j n_j \right].
\label{hamilton}
\end{equation}
Here $\Uhub$ is the Hubbard $U$-term giving the energy cost for 
double-occupation of a QD, and $\thub$ is the hopping parameter between 
adjacent QDs. In experiment $\thub$ can be set by regulating 
the height of the inter-dot tunneling barriers \cite{waugh}, and $\Uhub$
by adjusting the size of the QDs, since $\Uhub \sim 1/L$ for a QD
of size $L$.
We shall measure all energies in units of $\thub$, and set $\hbar = 1$.
The operators $c_{j \sigma} / c^{\dagger}_{j \sigma}$ are the 
electronic annihilation/creation operators, $n_{j \sigma}$ is the 
usual number operator, and $n_j = n_{j \uparrow} + n_{j \downarrow}$. 
The inter-dot spacing is denoted by $a$, and for 
simplicity we shall describe the time-dependent
electric field in terms of the potential difference between 
neighboring sites, $E(t) = e a F(t)$, which we parametrize as:
\begin{equation}
E(t) = E \sin \omega t, \quad T = 2 \pi / \omega .
\label{field}
\end{equation}

Experimental investigations of semiconductor QDs have revealed that
typically the spin relaxation time within these structures
is very long \cite{fuji} and accordingly we do not include any spin-flip 
terms in the Hamiltonian (\ref{hamilton}). As a result
the singlet and triplet sub-spaces of the model are completely decoupled.
In this work we consider just the singlet sub-space, as in
the triplet sub-space the Pauli principle forbids double-occupation
of a QD, and the Coulomb interaction described by the Hubbard-$U$
term is consequently irrelevant.
Suitable basis states for the spatial component of the two-electron
wavefunction can be obtained by taking symmetric combinations
of single-particle states defined on the lattice sites
$m$ and $n$ :
\begin{eqnarray}
| m, n \rangle &=& \frac{\phi_m(r_1) \phi_n(r_2) +
\phi_n(r_1) \phi_m(r_2)}{\sqrt{2}}, \ m \neq n 
\label{single} \\
| m, n \rangle &=& \phi_m(r_1) \phi_n(r_2), \qquad m = n
\label{double}
\end{eqnarray}
where $r_1, r_2$ are the 
coordinates of the two electrons. It can be clearly seen that
the basis consists of $N (N-1)/2$ entangled states (Eq.\ref{single})
in which the two electrons occupy different QDs, with the remaining $N$ 
factorisable states (Eq.\ref{double}) being those in which a QD is 
doubly-occupied. For strong values of the Coulomb interaction the energies 
of these two types of states will fall into two bands, separated 
approximately by the Hubbard gap $\Uhub$.

Our numerical investigation consists of initializing the system in
a given state, and propagating it forward in time under the action of
the time-dependent Hamiltonian (\ref{hamilton}) over many (typically
between ten and one hundred) cycles of the driving field. This can be 
done efficiently by a fourth-order Runge-Kutta method.
To study its time evolution quantitatively, 
we measure the projection of the two-electron wavefunction onto the 
basis states 
\begin{equation}
P_{m n} (t) = \left| \langle \Psi(t) | m, n \rangle \right|^2  
\label{project}
\end{equation}
together with the distribution of electronic charge within the array, 
and at all times we check the normalization of the wavefunction to
verify that unitarity is adequately preserved by the numerical procedure.

As the Hamiltonian (\ref{hamilton}) is periodic in time, $H(t) = H(t+T)$,
the Floquet theorem allows us to write solutions of the time-dependent
Schr\"odinger equation as $|\Psi_j (t) \rangle = \exp[-i \epsilon_j t] 
|\phi_j (t) \rangle$, where $|\phi_j (t) \rangle$ are termed Floquet states
and $\epsilon_j$ are the Floquet quasienergies. The Floquet states have
the same periodicity as the Hamiltonian, and are eigenvectors
of the unitary time-evolution operator for one period of the driving:
\begin{equation}
U(t_0+T,t_0) = {\cal T} \exp \left[-i \int_{t_0}^{t_0+T} H(t') dt' 
\right] ,
\label{unitary}
\end{equation}
where ${\cal T}$ signifies time-ordering. The corresponding
eigenvalues of this operator are related to the quasienergies 
\cite{fbz} via $\lambda_j = \exp[-i \epsilon_j T]$. As the Floquet states
provide a complete basis, the two-electron wavefunction
can be expanded as:
\begin{equation}
|\Psi(t) \rangle = \sum \ \alpha_j \ \exp[-i \epsilon_j t] \
|\phi_j (t) \rangle .
\label{expand}
\end{equation}
This expansion is directly analogous to the expansion of a wavefunction
in energy eigenstates in the case of a time-independent Hamiltonian,
and indeed in the adiabatic limit $T \rightarrow \infty$ the quasienergies
evolve into eigenenergies, and the Floquet states to eigenstates.  
From Eq.\ref{expand} it is clear that the long timescale behavior of
the system is dictated essentially by just the quasienergies, since the 
Floquet states are periodic with period $T$, and also generally have a weak 
time-dependence. Tunneling between two states is suppressed 
\cite{hanggi_rep} when their 
quasienergies approach each other as an external parameter (such as the 
electric field) is varied. Thus as well as being directly observable in 
the behavior of the electronic wavefunction, CDT may also be detected in 
the system's quasienergy spectrum by the presence of exact or avoided 
crossings. We shall show in the next section how these crossings 
in the exact quasienergies, obtained from the numerical diagonalization 
of the time-evolution operator (\ref{unitary}), correspond to the 
suppression of the dynamics of the electrons, and how analytic approximations
to the quasienergies allow the parameters at which this occur to be
predicted with high accuracy.

\section{Results}

In Fig.\ref{quasi} we show the exact quasienergy spectra with parameters 
$\omega = 2$ and $\Uhub = 16$, for two extreme cases of array size: 
a two-site system (i.e. a double QD) and a large system of $N=16$ 
sites. To assist the comparison between the two systems, the quasienergies
have been classified into three categories according to how
the electrons are distributed within the QD array in the corresponding
Floquet state. To determine this
we first class the basis states as 
{\it i)} states in which both electrons occupy the same QD 
(``doubly-occupied states''), 
{\it ii)} states in which the electrons occupy neighboring QDs 
(``neighbor-states'') and {\it iii)} states in which the
electrons are more widely-separated. We then assign each Floquet state to
the class onto which it projects the most, according to the magnitude
of the overlap
integrals $\left| \langle \phi_j(0) | m ,n \rangle \right|^2$. 

\begin{widetext}
\begin{center}
\begin{figure}
\includegraphics[width=.9\textwidth]{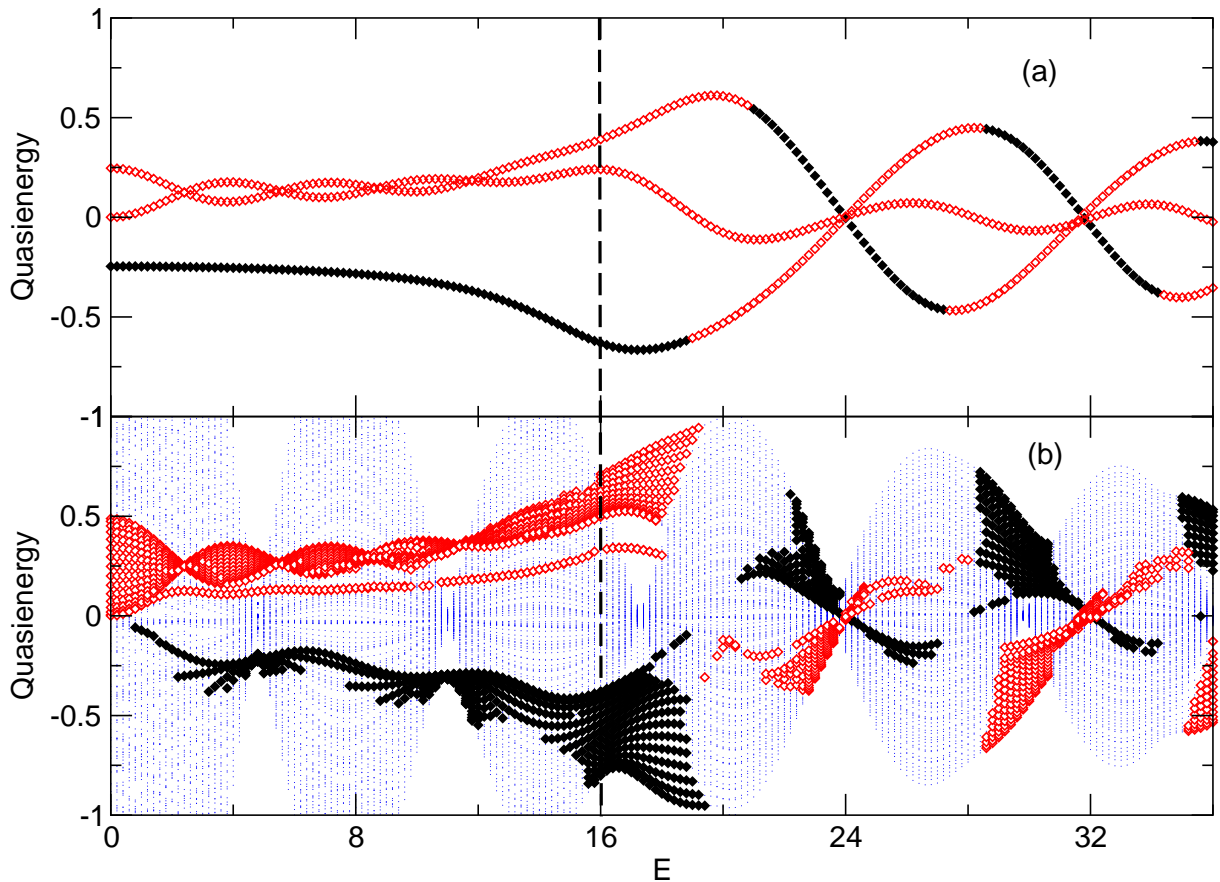}
\caption{(Color online) Quasienergy spectrum for (a) a two-site system,
and (b) a 16-site system, for $\Uhub=16$ and $\omega=2$.
Symbols indicate the characteristic of the corresponding Floquet state:
hollow diamonds (red) = doubly-occupied states, 
solid diamonds (black) = neighbor-states, 
points (blue) = states with wide separation between electrons.
The widely-separated states (points) exhibit a series of miniband 
collapses in excellent
agreement with Eq.\ref{pert_nonint} over the whole range
of $E$. For weak driving the doubly-occupied states show
a similar set of collapses but with {\em half} the period.
The vertical dashed line marks the boundary between
the weak and strong driving regimes.}
\label{quasi}
\end{figure}
\end{center}
\end{widetext}

Some similarities between the two spectra in Fig.\ref{quasi} are immediately 
clear. In the two-site case a pair of quasienergies make a sequence of 
exact crossings as $E$ is increased from zero up to $\Uhub$, 
while the remaining quasienergy stays isolated from them. 
As the two inter-crossing states 
evolve from the excited eigenstates of the static Hamiltonian, 
they consequently project mainly onto the 
doubly-occupied states. The remaining, isolated, 
Floquet state evolves from the system's ground state
and therefore has the character of a neighbor-state.

\begin{center}
\begin{figure}
\includegraphics[width=.45\textwidth]{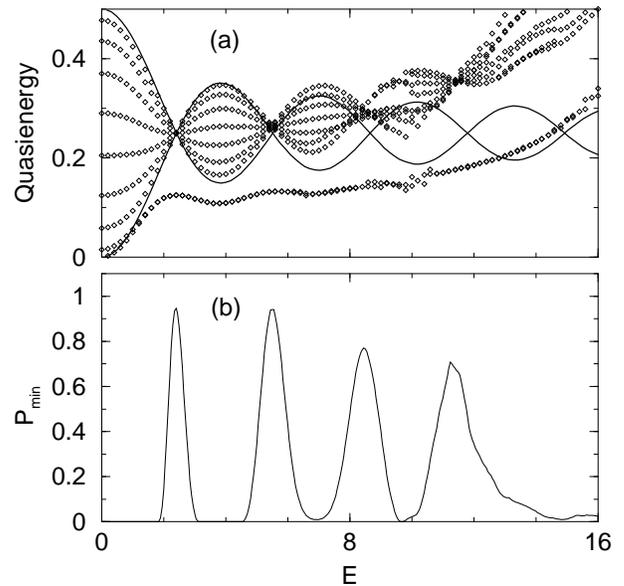}
\caption{Weak-field regime for a nine-site system, $\Uhub=16$ and $\omega=2$. 
(a) Seven quasienergies form a miniband, while the two almost degenerate 
edge-states remain isolated from the rest. The solid lines are the 
approximate solution $\pm (\Delta/2) J_0(2E/\omega)$, where $\Delta$ 
is the width of the miniband. As $E$ approaches $\Uhub$ the miniband
structure progressively breaks down.
(b) Minimum value of $P_{55}(t)$ for this system, 
initialized in state $|5,5\rangle$. At the points of quasienergy degeneracy 
(miniband collapse) CDT occurs.}
\label{weak}
\end{figure}
\end{center}

In the $16$-site system we can see a similar, but richer, behavior. 
Instead of just two inter-crossing states in the weak-field regime, 
the set of doubly-occupied states (\ref{double})
form a {\em miniband}, similar to that observed for non-interacting 
electrons in Refs.\cite{holthaus,holt_zpb}. In analogy to the crossings seen 
in the two-site system, this miniband
exhibits a sequence of ``band-collapses'' at which the quasienergies
become degenerate. These collapses occur at the same values of $E$ at 
which the crossings occur in the two-site case.
This behavior is seen more clearly in Fig.\ref{weak}a 
for a nine-site system. Here seven of the doubly-occupied quasienergies make 
up a miniband and exhibit a set of band-collapses as $E$ is increased, while 
the two remaining states, which physically correspond to the two edge-states, 
are almost degenerate and show little dependence on $E$.

For {\em non-interacting} particles, Holthaus \cite{holt_zpb} 
obtained an approximate expression for the quasienergy spectrum:
\begin{equation}
\epsilon_j = \frac{\Delta}{2} \ J_0(E/\omega) \ \cos a k_j
\label{pert_nonint}
\end{equation}
where $J_0$ is the zeroth Bessel function of the first kind,
$\Delta$ is the width of the miniband, and $k_j$ are the permitted
lattice momenta (assuming periodic boundary
conditions). In the system we consider, the Coulomb interaction only
operates if both electrons occupy the same QD. Accordingly,
if the electrons are distant from each other they behave effectively as
free particles, and in Fig.\ref{quasi}b we can see that the Floquet
states in which the two electrons are widely separated (plotted with points)
form a miniband of exactly the type given by Eq.\ref{pert_nonint}. 
Clearly these states 
cannot arise in the two-site system, as it is impossible for the two 
electrons to be further apart than a single lattice spacing.

Surprisingly, Holthaus' result for non-interacting particles
can also be applied to understand
the miniband behavior of the doubly-occupied states, in which 
the Coulomb interaction is {\em maximized}.
In this case, as the miniband is composed of Floquet states
that project onto doubly-occupied states, it is plausible
to visualize the localized two-electron state as tunneling
from site to site as a single object. Since this object has charge
$2e$ the period of the miniband is consequently halved,
taking the form
$\epsilon_j = (\Delta / 2) \ J_0(2 E/\omega) \ \cos a k_j$.
In Fig.\ref{weak}a we plot the envelope of this miniband, and see that for
low values of $E$ the agreement with the exact results is excellent.
As the value of $E$ increases, however, the doubly-occupied Floquet 
states begin to mix and interact with the other states, and the miniband 
structure starts to break down. When $E$ exceeds the size of the Hubbard gap
the structure is completely lost, and the system enters the 
strong-driving regime.

To illustrate the physical significance of the miniband collapses, we
show in Fig.\ref{weak}b the minimum value of the probability
$P_{55}(t)$ attained by the system during
ten periods of the driving, having been initialized in the
state $|5, 5 \rangle$ -- that is, with both electrons in the central QD.
It can readily be seen that at the points of miniband collapse the
system remains frozen in its initial state, thereby manifesting CDT, and
producing sharp peaks in $P_{\mbox{\small min}}$ centered on the points of 
quasienergy degeneracy. This demonstrates that interpreting the two-electron
state as a single composite particle, as suggested by the form of the
Floquet spectrum, indeed provides the correct description of the system's 
dynamics.

\begin{center}
\begin{figure}
\includegraphics[width=.45\textwidth]{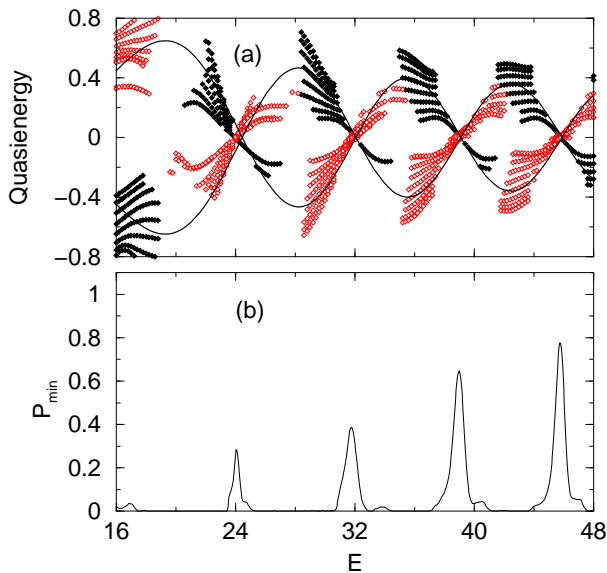}
\caption{(Color online) 
Strong field regime for the nine-site system. (a) Crossings occur
between Floquet states of doubly-occupied type
(hollow symbols (red)) and those of neighbor type (solid symbols (black)).
The lines are the perturbative solution $\pm 2 J_8(E/\omega)$.
(b) Minimum value of $P_{55}(t)$ for this system, initialized in
state $|5,5\rangle$. CDT again occurs when the quasienergies cross.}
\label{strong}
\end{figure}
\end{center}

To complete the comparison of the two cases in Fig.\ref{quasi}, we now 
consider the strong-driving regime ($E > \Uhub$). Again, clear similarities 
between the two quasienergy spectra are evident. In the case of the two-site 
system, a pair of the quasienergies make a set of close approaches while the 
third oscillates weakly around zero. Detailed examination of the close 
approaches (Ref. \cite{cecgp1}) reveals that they are in fact
{\em avoided crossings}, and a perturbative treatment of this system 
performed in that work yields the approximate solution:
\begin{equation}
\epsilon_{\pm} = \pm 2 J_n(E/\omega) ,
\label{pert_int}
\end{equation}
where $n \omega = \Uhub$. When this resonance condition between $\Uhub$
and the driving frequency is not satisfied, the quasienergies in this
regime show little structure, and CDT is in general not produced.
In Fig.\ref{quasi}b it is clear that a generalization of this
phenomenon occurs, and a similar set of
close approaches between neighbor and doubly-occupied Floquet
states occurs. In
Fig.\ref{strong}a we plot the quasienergies of the neighbor and
doubly-occupied Floquet states for a nine-site array in this
regime, and find that the locations of the quasienergy degeneracies are
again given accurately by the perturbative result (Eq.\ref{pert_int}),
namely at the roots of
$J_8(E/\omega)$. It can also be seen that the agreement between the 
perturbative approximation and the exact result improves as $E$ increases.
This is to be expected, as the ``perturbation'' used in the derivation 
of Eq.\ref{pert_int} consists of the tunneling component of the Hamiltonian, 
and thus the result becomes exact in the limit of large $E$
when the tunneling is negligible in comparison with the ac-driving.

In Fig.\ref{strong}b we confirm that the structure seen in the
quasienergy spectrum indeed corresponds to a modification of the system's
dynamics. As before, the system is initialized in state $|5,5\rangle$, 
and integrated over ten cycles of the driving field. The sharp peaks
again visible in $P_{\mbox{\small min}}$
are located at the points of quasienergy degeneracy,
indicating that CDT occurs as expected.
Thus in both weak and strong driving regimes, Floquet analysis can
be used to predict the field parameters at which CDT will occur 
in a driven QD array with high accuracy.

\begin{center}
\begin{figure}
\includegraphics[width=.45\textwidth]{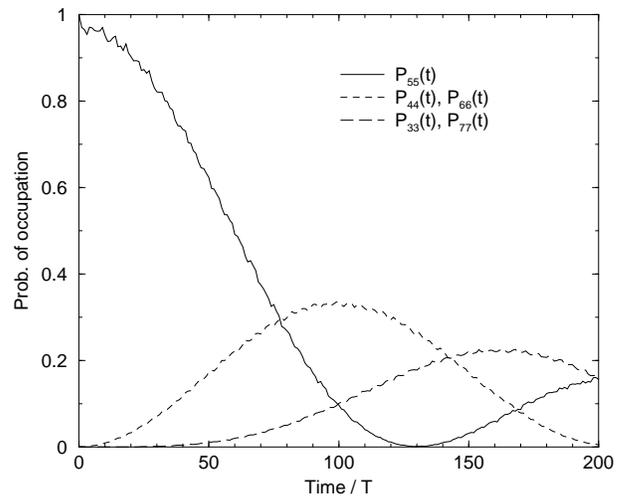}
\caption{Time-dependent projection of the two-electron
state onto the basis functions (see Eq.\ref{project})
with the system initialized in state $|5,5\rangle$.
Only the projections onto the doubly-occupied basis states
are shown, as all the others are negligible in size.
Field parameters: $E = 2.35$, $\omega = 2$. The differences between
$P_{44}$ and $P_{66}$, and between $P_{33}$ and $P_{77}$,
are too small to be seen here.}
\label{occupy}
\end{figure}
\end{center}

We now examine the dynamics of the system in the weak-driving regime
in more detail. If the system is initialized in a state in which
both electrons occupy the same QD, one would expect that the
Coulomb repulsion would act to repel the two electrons away from each
other. We have already seen, however, that this does not occur, and that
in the presence of the ac-field the two electrons in fact bind together
and propagate as a single composite particle \cite{cecgp3, noba}. 
Although surprising, the ability of an ac-field to convert a
repulsive interaction into a time-averaged attraction forms the
basis of an increasing number of applications, ranging from the 
well-known Paul trap \cite{paul} to a recent proposal to bind atomic nuclei 
together \cite{smirnova} with laser fields. We show in Fig.\ref{occupy}
the explicit time-dependence of the dominant components of the wave-function
of a nine-site system, initialized with both electrons in the central QD. 
The strength of the electric field has been set to $E = 2.35$, close 
to the point of the first miniband collapse at 
$E \simeq 2.4048$, and thus the effective
tunneling is heavily suppressed but non-zero.
It can be seen that over the first hundred periods of the driving
the initial state $|5, 5 \rangle$ evolves chiefly
into a superposition of the states $|4, 4 \rangle$ and $|6, 6 \rangle$, 
by diffusing symmetrically \cite{symm} to the neighboring QDs, while
retaining the two-particle superposition. 
This is confirmed by the fact that throughout the time-evolution,
the total projection of the wavefunction
onto the doubly-occupied states never takes a value of less than $0.955$,
indicating that the role of the other states is merely to act as 
short-lived intermediates in the diffusion process.

\begin{center}
\begin{figure}
\includegraphics[width=.40\textwidth]{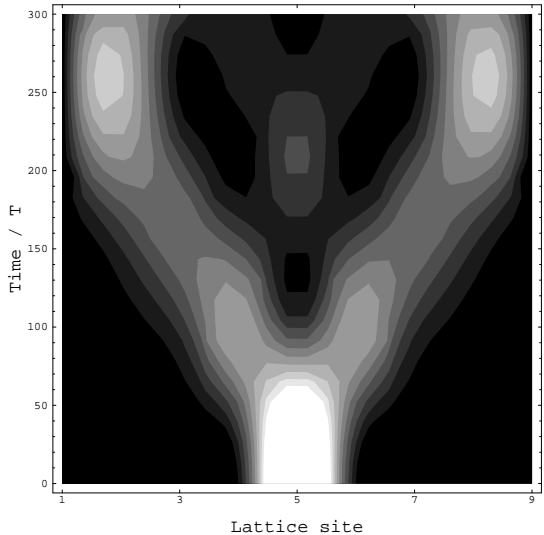}
\caption{Time-dependent charge-density, $|\Psi(t)|^2$, for a nine-site lattice, 
with the initial condition of both particles on site 5 (the central site).
White / black  signifies high / low values of density.
Field parameters are $E = 2.35$, $\omega = 2$.}
\label{density}
\end{figure}
\end{center}

\begin{center}
\begin{figure}
\includegraphics[width=.45\textwidth]{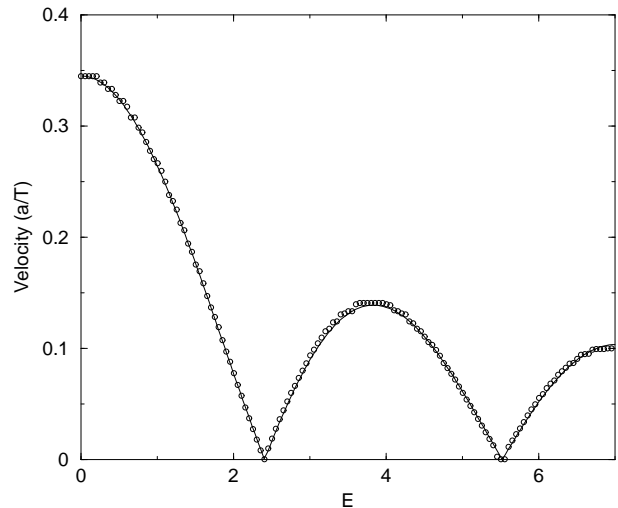}
\caption{Circles indicate the velocity of the charge pulse in the nine-site
system as it propagates through the array from the initial state
$|5, 5\rangle$. The solid line is \ $v_0 |J_0(2E/\omega)|$ where
$v_0$ is a fitting parameter.}
\label{velocity}
\end{figure}
\end{center}

Fig.\ref{density} vividly demonstrates how the initially
localized two-electron state on site $5$ splits smoothly into two 
pulses moving to the left and right at an almost constant 
velocity. Although reminiscent of soliton dynamics, the lack of any
dispersive effects to provide ``refocusing'' of the wavefront means that
as the pulses traverse the system they do smear out 
to an extent. A result of this, for example, is that the peak in $P_{33}$ 
does not rise to the same height as $P_{44}$ in Fig.\ref{occupy} 
due to blurring of the pulse-shape. Even so, the definition of
the charge pulses is sufficiently crisp to suggest the possibility
of using an array of QDs as a type of 
coherent electron turnstile \cite{tamb_prl}.
In such a device the ac-field is used to control the rate
at which the charge-pulses propagate, so that in a given interval of time
a specified quantity of charge is transferred along the array.
The well-defined trajectories of the current pulses visible in 
Fig.\ref{density} indicate that the
velocity of the pulse can be unambiguously measured, and
in Fig.\ref{velocity} we give a plot of this quantity, obtained as the
reciprocal of the time required for the initial QD to become empty.
Since the velocity of the pulse is directly related to the magnitude
of the inter-dot tunneling, the velocity shows the same Bessel function
dependence as the quasienergies, and vanishes at field parameters
such that $J_0 (2 E /\omega) = 0$, at which the tunneling is completely 
suppressed. This clearly shows how the ac-field can be used 
as a control parameter to regulate the speed of propagation of the pulse by 
means of CDT.

It is interesting to note that this arrangement 
generates an unusual current of entangled, correlated electrons.
As the initial state propagates symmetrically to the left and right
of the QD array, there is an equal probability of finding
an individual electron on either side of the central QD, but due to the
combined effect of the ac-field and the Coulomb interaction there is a
very high chance that the other electron will also be occupying the same QD. 
Recently Saraga and Loss \cite{saraga} have proposed using an
array of three QDs to create currents of entangled electrons
for use in quantum computing and communication applications.
We note that their setup contains many similarities with the system 
we study here: a pair of electrons is injected into the central QD,
which then emits electron pulses to its neighbors.
The major difference is that in Ref.\cite{saraga} the current is
composed of single electrons (in a superposition of up and down 
spin states), whereas in our case the current
is carried by a two-electron bound state. 
In this sense the driven QD array that we study
can be regarded as an electronic analogue
of an optical beam-splitter \cite{oliver},
which divides the initial state into two 
spatially separated and entangled electron pairs moving
at a velocity controlled by the ac-driving field.

\section{Conclusions}

In summary, we have studied the dynamical behavior of two interacting 
electrons in an QD array under ac-driving. We have seen that the competition 
between the Coulomb energy and the driving field produces a rich 
phenomenology, which can be described well by means of Floquet theory. 
Throughout the weak field regime, the driving field and the Coulomb interaction 
combine in a counter-intuitive way to bind the two electrons together
to form a composite particle resembling an exciton.
The tunneling of this particle from site to site can then
be controlled by manipulating 
the driving field, which permits either the complete freezing
of the dynamics at the points of ``miniband collapse'', or the 
fine regulation of the current of entangled electron pairs. 
When the field is stronger than the Coulomb
interaction the character of the Floquet states abruptly changes, 
and a different form of CDT occurs, arising from the close approaches
of states of doubly-occupied type and neighbor-type. This
generalizes the form of CDT observed previously in
a two-site system \cite{cecgp1}, and only occurs when the driving
field is in resonance with the Coulomb energy (i.e. $n \omega = \Uhub$)
at the roots of $J_n(E/\omega)$. The delicate control over
the electron dynamics afforded by these effects
has many potential applications to the coherent manipulation of
correlated electron states in mesoscopic systems. 

Although in this work we only consider one specific set of system 
parameters and one driving frequency, we have found that the results are 
of general validity as long as $\Uhub > \thub$ (which is usually satisfied 
for QDs in the Coulomb blockade regime) and that the system is in the 
high-frequency regime \cite{hanggi_epl} ($\omega > \thub$). 
A recent experimental study \cite{elzerman} gave estimates of 
$\Uhub \simeq 3.7$ meV and $\thub \simeq 0.5$ meV for the double QD system 
studied there. Given that the QD separation was about 200 nm, the required 
driving field to see the CDT effects we report should be of the order of 
200 GHz, with field-strengths of up to 10 kV/cm. Such systems also
possess excellent coherence properties, with decoherence times of
the order of nanoseconds reported recently \cite{hayashi}.
We therefore believe 
that experimental observation of these effects is feasible, and could
be achieved, for example, by connecting the system to external leads and
measuring the current passing through the array, or more directly,
by measuring the charge occupation of the individual QDs by means of
quantum point contacts, as has recently been achieved in 
Refs.\cite{elzerman, lu}. This latter technique has the important
advantages that the occupation of the QDs can be determined even if the 
inter-dot current is too small to measure by conventional means, and also
should induce less decoherence than the transport form of measurement.

\bigskip

CEC acknowledges support from the 
physics department of the Universit\`a di Roma ``La Sapienza''.
GP was supported by the Spanish DGES grant MAT2002-02465 and by the
EU Human Potential Programme under contract HPRN-CT-2000-00144.


\begin{thebibliography}{99}
\bibitem{fbz}
{The quasienergies are only defined up to
an arbitrary multiple of the driving frequency, that is
$\epsilon$ and $\epsilon + n \omega$ correspond to the same 
physical state. In this work we take the normal convention of
restricting the quasienergies to the ``first Brillouin zone''
($-\omega/2 \leq \epsilon < \omega/2$) which gives a unique representation.}
\bibitem{symm}
{In fact a small asymmetry between diffusion to the left and to the 
right is introduced by the phase of the driving field (Eq.\ref{field}),
but after averaging over many periods of the driving this 
asymmetry becomes negligible.}
\end{thebibliography}
\end{document}